\documentclass[11pt]{article}
\usepackage{geometry}                % See geometry.pdf to learn the layout options. There are lots.
\usepackage{graphicx,xcolor,hyperref}
\usepackage{amssymb}
\usepackage{epstopdf}
\usepackage{natbib}
\usepackage[T1]{fontenc}
\DeclareRobustCommand{\VAN}[3]{#2}
\let\VANthebibliography\thebibliography
\def\thebibliography{\DeclareRobustCommand{\VAN}[3]{##3}\VANthebibliography}
\usepackage{graphicx}	% Including figure files
\usepackage{amsmath}	% Advanced maths commands
\usepackage{amssymb}	% Extra maths symbols
\usepackage{bm} % boldfaced expression
\usepackage{tabularx}

\newcommand{\lcdm}{$\Lambda$CDM}
\newcommand{\de}{\mathrm{d}}
\newcommand{\cosmosis}{\texttt{CosmoSIS}}
\newcommand{\emcee}{\texttt{emcee}}
\newcommand{\multinest}{\texttt{multinest}}
\newcommand{\camb}{\texttt{CAMB}}

\title{Towards simulating a realistic data analysis with an optimised angular power spectrum of spectroscopic galaxy surveys}
\author{\textsc{
Guglielmo Faggioli,$^1$ %\thanks{E-mail: guglielmo.faggioli@edu.unito.it}
Konstantinos Tanidis,$^{1,2}$ %\thanks{E-mail: tanidis@to.infn.it}
\& Stefano Camera$^{1,2,3}$%\thanks{E-mail:stefano.camera@unito.it}
}}
\date{\small
$^{1}$Dipartimento di Fisica, Universit\`a degli Studi di Torino, via P.\ Giuria 1, 10125 Torino, Italy
\\$^{2}$INFN, Sezione di Torino, via P.\ Giuria 1, 10125 Torino, Italy\\
$^{3}$INAF, Osservatorio Astrofisico di Torino, strada Osservatorio 20, 10025 Pino Torinese, Italy
}                                           % Activate to display a given date or no date

\begin{document}
\maketitle

\begin{abstract}
The angular power spectrum is a natural tool to analyse the observed galaxy number count fluctuations. In a standard analysis, the angular galaxy distribution is sliced into concentric redshift bins and all correlations of its harmonic coefficients between bin pairs are considered---a procedure referred to as `tomography'. However, the unparalleled quality of data from oncoming spectroscopic galaxy surveys for cosmology will render this method computationally unfeasible, given the increasing number of bins. Here, we put to test against synthetic data a novel method proposed in a previous study to save computational time. According to this method, the whole galaxy redshift distribution is subdivided into thick bins, neglecting the cross-bin correlations among them; each of the thick bin is, however, further subdivided into thinner bins, considering in this case all the cross-bin correlations. We create a simulated data set that we then analyse in a Bayesian framework. We confirm that the newly proposed method saves computational time and gives results that surpass those of the standard approach.
\end{abstract}

\section{Introduction}
The forthcoming generation of experiments targeting the large-scale cosmic structure will provide us with data of exquisite quality, from which it will be possible to extract cosmological information to test the our current cosmological model (\lcdm), for instance investigating the nature of dark energy and dark matter. The two main probes envisaged for such experiments are weak gravitational lensing and galaxy clustering. In this paper, we shall focus on the latter.

Forthcoming galaxy surveys, such as the \textit{Euclid} satellite, \citep{Laureijs2011,Amendola2013,Amendola2016}, the Legacy Survey of Space and Time \citep{2009arXiv0912.0201L}, and the Square Kilometre Array \citep{SKA_paper}, will be characterised by a high computational time cost in their analysis, motivating the search for new optimised methods. For this reason, this work aims at developing an improved analysis technique, taking inspiration from \citet{stef_specz}. In particular, we adopt the philosophy there presented and implement it in a likelihood-based approach, simulating a synthetic data set that we then fit against the theoretical model predictions.

This paper is outlined as follows. In \autoref{sec:methods}, we introduce the survey assumptions considered throughout our analysis, present the harmonic-space angular power spectrum for galaxy clustering, describe in detail the optimised method, and show the likelihood and the scale cuts applied for the analysis. In \autoref{sec:results}, we discuss the results obtained with the standard and the optimised method. Finally, conclusions are presented in \autoref{sec:conclusions}.

\section{Methods}
\label{sec:methods}
\subsection{Survey assumptions}
We adopt the same survey specifications of \citet[][see their Section~2.2 and references therein, for details]{stef_specz}, who first proposed the method and tested it via a Fisher matrix analysis. Specifically, we consider a spectroscopic galaxy survey targeting H$\alpha$ emitters in the redshift range between $0.6$ and $2$, with an accuracy that can be modelled with a redshift-dependent Gaussian uncertainty on the distribution on the measured redshift with width $\sigma_z=0.001(1+z)$. The linear galaxy bias is modelled as $b(z)=\sqrt{(1+z)}$.

\subsection{The harmonic-space galaxy power spectrum} \label{ref:harmonic_space}
The harmonic-space (also, angular) power spectrum represents the natural tool to probe fluctuations in the observed galaxy distribution as measured from our point of view as observers. For large multipole values, $\ell \gg 1$, it is possible to employ the Limber approximation \citep{1953ApJ...117..134L,1992ApJ...388..272K}, to reduce the computational effort thanks to its collapsing a three-dimensional integral into a one-dimensional one. Under this assumption, the theoretical power spectrum of galaxy number counts for the redshift bin pair $i-j$ and on linear scales reads
\begin{equation}
    C_\ell^{ij}=\int\frac{\de\chi}{\chi^2}\,W^i(\chi)W^j(\chi)P_{\rm lin}\left(k=\frac{\ell+1/2}{\chi}\right),
    \label{eq:signal}
\end{equation}
where $\chi(z)$ is the comoving distance to redshift $z$,
\begin{equation}
    W^i(\chi)=n^i(\chi)b(\chi)D(\chi)
    \label{eq:kernel}
\end{equation}
is the window function in the $i$th redshift bin, with $n^i(\chi)$ its normalised galaxy distribution, $b(\chi)$ is the linear galaxy bias, and $D(\chi)$ the linear growth factor. Finally, $P_{\rm lin}(k)$ is the linear matter power spectrum at $z=0$, which is here provided by the Boltzmann solver \camb\ \citep{LCL2000}.

It is worth noting that the observed clustering of galaxies contains other terms on top of what we have described above, which is due to perturbations in the underlying matter density distribution \citep{BonvinDurrer2011,ChallinorandLewis2011}. The most notable of such terms are redshift-space distortions (RSD) and lensing magnification.\footnote{For a window function accounting for other terms on top of the density field in the Limber approximation we refer the reader to the relevant literature \citep[see][]{Tanidis:2019mag,Chisari_2019}.} However, these terms are suppressed on the scales of interest in this analysis and for the bin sizes we adopt, meaning we can safely neglect to include them.

\subsection{The traditional approach}
\label{sec:Std}
On the one hand, data from spectroscopic galaxy surveys has customarily been analysed in terms of the Fourier-space power spectrum and its decomposition into Legendre multipoles. Whilst this approach has worked perfectly, for the redshift and sky coverages of data hitherto collected, it is arguable that some of its underlying assumptions
%(e.g.\ the flat-sky approximation) 
will no longer be met with the next generation of cosmological experiments \citep[see e.g.][]{Ruggeri_2018,Blake_2018}. Moreover, the fluctuations in the observed galaxy number counts contain terms that cannot be decomposed in Fourier modes, like the nonlocal contribution from gravitational lensing, which will be all the more important for deeper surveys \citep{2015aska.confE..25C,PhysRevD.94.043007}.

On the other hand, the standard tomographic approach for the computation of the galaxy angular power spectrum $C^{ij}_{\ell}$ is based on all correlations among bin pairs $i-j$ across the whole redshift range. Now, the benchmark survey described in Section 2.1 will easily be able to slice the observed galaxy distribution in bins of width $\sim0.01$, which, for the redshift range considered, results into about $10^4$ between auto- and cross-bin correlations. Such a number has to be further multiplied by the number of bins in multipole space the data will be binned into. This is clearly computationally unfeasible, in the prospect of a likelihood-based analysis scanning---at the very least---the six-dimensional parameter space of the `vanilla' \lcdm\ model.

\subsection{The new method}
\label{sec:Hybrid}
This conundrum motivates the research of new strategies to analyse forthcoming surveys data sets. Among the different proposals, we follow \cite{stef_specz}, who proposed to combine relevant aspects of the two standard techniques described above. Fourier-space analyses usually employ a thick redshift binning, e.g.\ with width $\Delta z \approx 0.1$; all Fourier modes inside the bin are then considered, but the correlations among adjacent $z$-bins are not taken into account. However, applying this approach face-value to the harmonic-space $C^{ij}_{\ell}$ means losing information by squashing all the galaxies within the relatively large $\Delta z$ bin onto a single redshift slice.

Hence, the idea is to combine the two approaches in a `hybrid' method. This method is characterised by two binning tiers: the galaxy distribution is binned by adopting a set of top-hat \textit{thick} bins; each of these is further binned into top-hat \textit{thin} bins, convolved with a Gaussian in order to take into account for the small although non-negligible errors in the spectroscopic redshift estimation. This division is made by using equi-spaced bins. Each thick bin is considered as an independent survey, hence cross-correlation between them is not computed, while it is for the thin bins inside the thick ones. The resulting tomographic matrix $C^{ij}_{\ell}$ is thus block diagonal by construction.

\begin{figure}
\centering 
\includegraphics[width=1.0\textwidth]{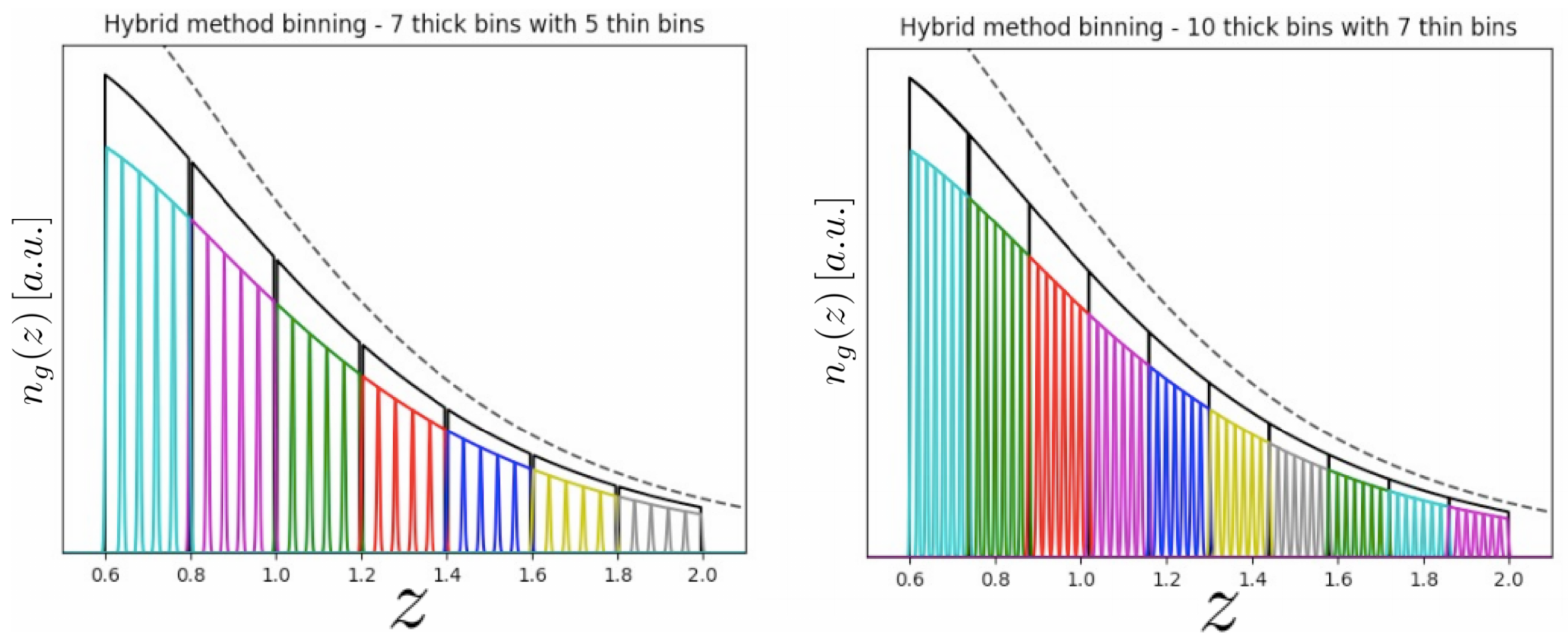}
\caption{.The top dashed black curve shows the unbinned galaxy distribution, $n_{\rm g} (z)$. Black curves correspond to thick bins, whilst coloured ones to thin bins inside each thick bin. (To enhance readability we have rescaled all the distributions by arbitrary factors.)}
    \label{fig:bins}
\end{figure}

In this paper, we include two hybrid binning configurations in the same redshift range $z\in[0.6,2.0]$, both smoothed by a Gaussian with $\sigma_z = 0.001$:
\begin{enumerate}
    \item 7 equi-spaced thick bins of redshift width $\Delta z=0.2$, each having 5 equi-spaced thin bins of width $\delta z=0.04$. This case is represented by black and coloured bins, respectively, in the left panel of \autoref{fig:bins};
    \item 10 equi-spaced thick bins of redshift width $\Delta z=0.14$, each having 7 equi-spaced thin bins of width $\delta z=0.02$, as shown in the right panel of \autoref{fig:bins}.
\end{enumerate}

\subsection{Set-up of statistical analysis}
To construct the likelihood and forecast constraints on the cosmological parameters of interest, we employ the publicly available suite \cosmosis\ \citep{Zuntz2015}, which we modify to reproduce the hybrid method described above. We generate our synthetic data vector by using the angular power spectra as given in \autoref{eq:signal}. We choose as a reference a flat \lcdm\ model with the cosmological parameter set $\bm \theta=\{\Omega_{\rm m},\,h,\,\Omega_{\rm b},\, n_{\rm s} ,\, \ln{(10^{10}A_{\rm s})} \}$, whose fiducial values are $\bm \theta_{\rm fid}=\{0.31,\,0.6774,\,0.05,\, 0.9667 ,\, 3.06 \}$. For details on the samplers and analysis employed to explore the parameter space, see \autoref{sec:results}. 

For the data, we assume a Gaussian likelihood, and we focus on minimising the chi-squared. In other words, we do not include the likelihood normalisation in the parameter estimation. This assumption does not hinder our result, as the data covariance is assumed independent of the cosmological parameters.

Concerning the covariance of the galaxy clustering signal given in \autoref{eq:signal}, we adopt the Gaussian approximation, namely
\begin{equation}
{\rm Cov}\left(C^{ij} _\ell,C^{mn} _{\ell^\prime}\right)=
\frac{\delta_{\rm K}^{\ell\ell^\prime}}{2\ell\Delta\ell f_{\rm sky}}\left[
\left(C^{im} _\ell +\frac{\delta^{im}_{\rm K}}{\bar n_i}\right)
\left(C^{jn} _\ell +\frac{\delta^{jn}_{\rm K}}{\bar n_j}\right)+
\left(C^{in} _\ell +\frac{\delta^{in}_{\rm K}}{\bar n_i}\right)
\left(C^{jm} _\ell +\frac{\delta^{jm}_{\rm K}}{\bar n_j}\right)
\right],
\label{eq:covmat}
\end{equation}
where $\Delta \ell$ is the multipole binning width, $f_{\rm sky}$ the sky fraction covered by the survey, $\delta_{\rm K}$ the Kronecker symbol and $\bar n_i$ is the surface galaxy density in bin $i$.

The angular power spectra are computed with the Limber approximation and in the linear regime, we therefore focus on multipole range, $\ell \in [100,800]$ as a reasonable interval. It is possible that for a few bin-pair configurations either the lower or upper multipole limit exceeds the range of validity of the Limber approximation or the nonlinear scale. However, we do not aim to make forecasts for a specific experiment but rather to compare the performance of the standard and hybrid methods in a realistic setting, and thus this choice does not affect our conclusions. In both binning scenario we consider $n_\ell=5$ log-spaced multipole values in the aforementioned range.

%Before we proceed with the presentation of our results, it is important to note that in the hybrid method, outlined already in \autoref{sec:Hybrid}, we treat each thick bin as an independent survey having its own likelihood evaluation. This is afterwards multiplied with the other thick bins independent likelihoods forming the joint likelihood under the same parameter value set provided by the sampler.  This is not obvious but can be easily implemented thanks to the modular properties of the \cosmosis\ package. 

\section{Results}
\label{sec:results}
Here, we present and compare the results obtained with the standard and the hybrid methods. As already mentioned in \autoref{sec:Hybrid}, we applied two hybrid binning configurations in the redshift range $z\in[0.6,2.0]$. We can summarise our findings as follows. All cosmological parameter reconstructed mean values and inferred $68\%$ confidence level intervals are summarised in \autoref{table:Standard method analyses results}.

\begin{enumerate}
    \item For the standard approach we use 20 equi-populated bins in the redshift range $z\in[0.6,2.0]$, and $n_{\ell}$ = 5 multipole values in the considered $\ell$ range. We use the \multinest\ sampler \citep{multinest_sampler} to forecast constraints.\\

    \item Regarding the first configuration of the hybrid binning we use equi-spaced thick bins with width $\Delta z = 0.2$, in the same redshift range, while for the thin bins we use a width $\delta z = 0.04$. This means that we have seven thick bins, considered as seven independent surveys, each of them containing five thin bins. Again, we use $n_{\ell}$ = 5 multipole values while for the sampling method we chose the \emcee\ sampler \citep{emcee_paper}, better suited for the tomographic matrix configuration of the hybrid method. \\
    
    \item In the second hybrid binning configuration the thick bin width is $\Delta z = 0.14$ and the thin bins $\delta z = 0.02$, working now with finer binning of 10 thick bins each containing 7 thin bins. The sampler employed is, again, \emcee.
\end{enumerate}
    \begin{table}
	    \centering
	    \caption{Summary of analysis results for each parameter (first column) with: its input fiducial value, $\theta_{\rm fid}$ (second column); reconstructed mean value, $\theta^\ast$ (third, fifth, and seventh column); and 68\% confidence level error interval,  $\sigma_{\theta}$ (fourth, sixth, and eighth column).}
    	\begin{tabularx}{\textwidth}{ l l l X l l l l l l }
   	        \hline
	      & & \multicolumn{2}{c}{Standard} && \multicolumn{2}{c}{Hybrid (config.\ 1)} && \multicolumn{2}{c}{Hybrid (config.\ 2)}\\
	       \cline{3-4}\cline{6-7}\cline{9-10}
	      Parameter & \multicolumn{1}{c}{$\theta_{\rm fid}$} & \multicolumn{1}{c}{$\theta^\ast$} & \multicolumn{1}{c}{$\sigma_{\theta}$} && \multicolumn{1}{c}{$\theta^\ast$} & \multicolumn{1}{c}{$\sigma_{\theta}$} && \multicolumn{1}{c}{$\theta^\ast$} & \multicolumn{1}{c}{$\sigma_{\theta}$} \\
   	        \hline
   	        \hline
	    $\Omega_{\rm m}$ & $0.31$ & $0.310001$ & $0.00015$ && \ \ $0.310002$ & $0.000074$ && $0.310001$ & $0.000023$ \\ 
	    \hline
    	    $h$ & $0.6774$ & $0.685$ & $0.028$ && $0.684$ & $0.026$ && $0.681$ & $0.019$ \\
	    \hline
    	    $\Omega_{\rm b}$ & $0.05$ & $0.0511$ & $0.0028$ && $0.0505$ & $0.0026$ && $0.0499$ & $0.0023$ \\
	    \hline
    	    $n_{\rm s}$ & $0.9667$ & $0.964$ & $0.014$ && $0.0964$ & $0.013$ && $0.965$ & $0.011$ \\
	    \hline
    	    $\ln(10^{10} A_{\rm s})$ & $3.06$ & $3.048$ & $0.017$ && $3.049$ & $0.015$ && $3.053$ & $0.012$ \\
	    \hline
    	\end{tabularx}
	    \label{table:Standard method analyses results}
    \end{table}

For a more thorough comparison of the two methods, in \autoref{table:computational_times} we also show the computation times running a fixed cosmology on a specific parameter value set for the standard and the hybrid method. For sake of comparison of running time test, we consider a third hybrid binning with 14 thick bins each containing 10 thin bins while keeping the same smearing with the previous cases. It can be clearly seen from \autoref{table:computational_times} that the larger the number of the bins, the more time we save by using the hybrid method with respect to the standard one.
\begin{table}
	\centering
	\caption{Comparison between standard and hybrid computation times for a fixed cosmology.}
    	\begin{tabular}{ c  c  c }
   	 \hline
    	No.\ of bins & Standard method running time [s] & Hybrid method running time [s]  \\ \hline
    	\rule{0pt}{13pt} 35 & 7 & 5  \\ \hline
    	\rule{0pt}{13pt} 70 & 14 & 8  \\ \hline
    	\rule{0pt}{13pt} 140 & 22 & 11 \\ \hline
    	\end{tabular}
	\label{table:computational_times}
\end{table}

\begin{figure}
\centering 
\includegraphics[width=0.6\textwidth]{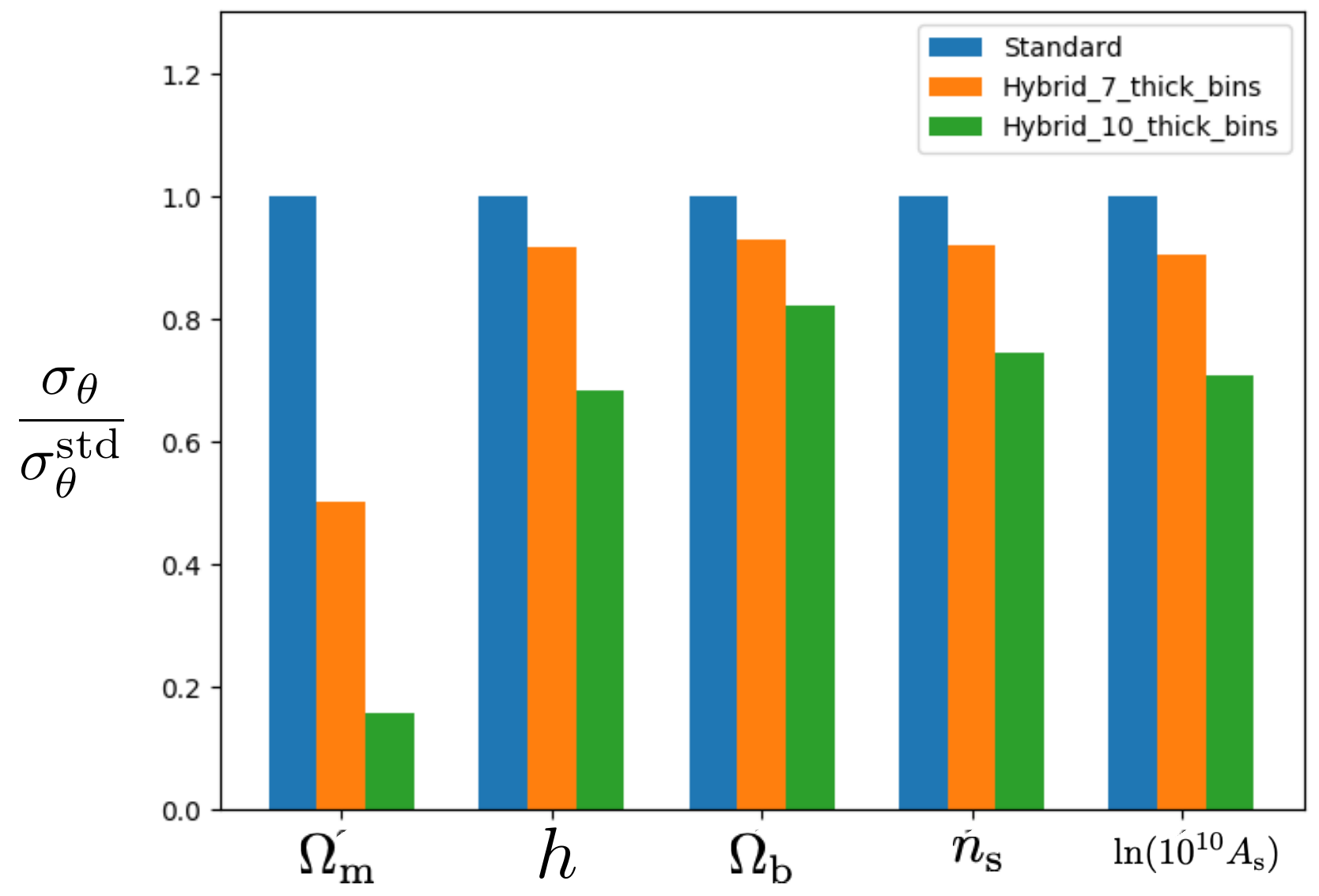}
\caption{Comparison between marginalised errors from the hybrid methd, $\sigma_{\theta}$, and what obtained from the standard approach, $\sigma_{\theta}^{\rm std}$, on the estimated cosmological parameters.}
    \label{fig:histo_3_configurations}
\end{figure}

Another major advantage of the hybrid method over the standard approach is that it yields tighter constraints on the parameter of interest. This is due to the fact that the finer binning of the thin bins allows us to recover partly the three-dimensional information encoded in the correlation of galaxies within the thick bin. To appreciate better the aforementioned enhancement in constraining power, in \autoref{fig:histo_3_configurations} we show the ratio between the $68\%$ marginal error intervals on each parameter from the hybrid method and the same obtained with the standard approach, for the two binning configurations of \autoref{sec:Hybrid} (green and red candlesticks, for configuration 1 and 2 respectively). Note that the blue candlesticks are the ratio of the $68\%$ marginal error intervals of the standard approach with themselves, simply to guide the reader's eye. This clearly shows us how the finer the binning, the tighter the constraints, both because we can track better the cosmic growth and the redshift evolution of the source distribution (thick binning), and because we can recover radial information (thin binning). Actually, the fact that even the seven thick bins of the hybrid binning configuration no.\ 1 perform better than the 20 bins of the standard method is a proof that radial information within the bin is crucial for accurate cosmological parameter estimation.

\begin{figure}
\centering 
\includegraphics[width=\textwidth]{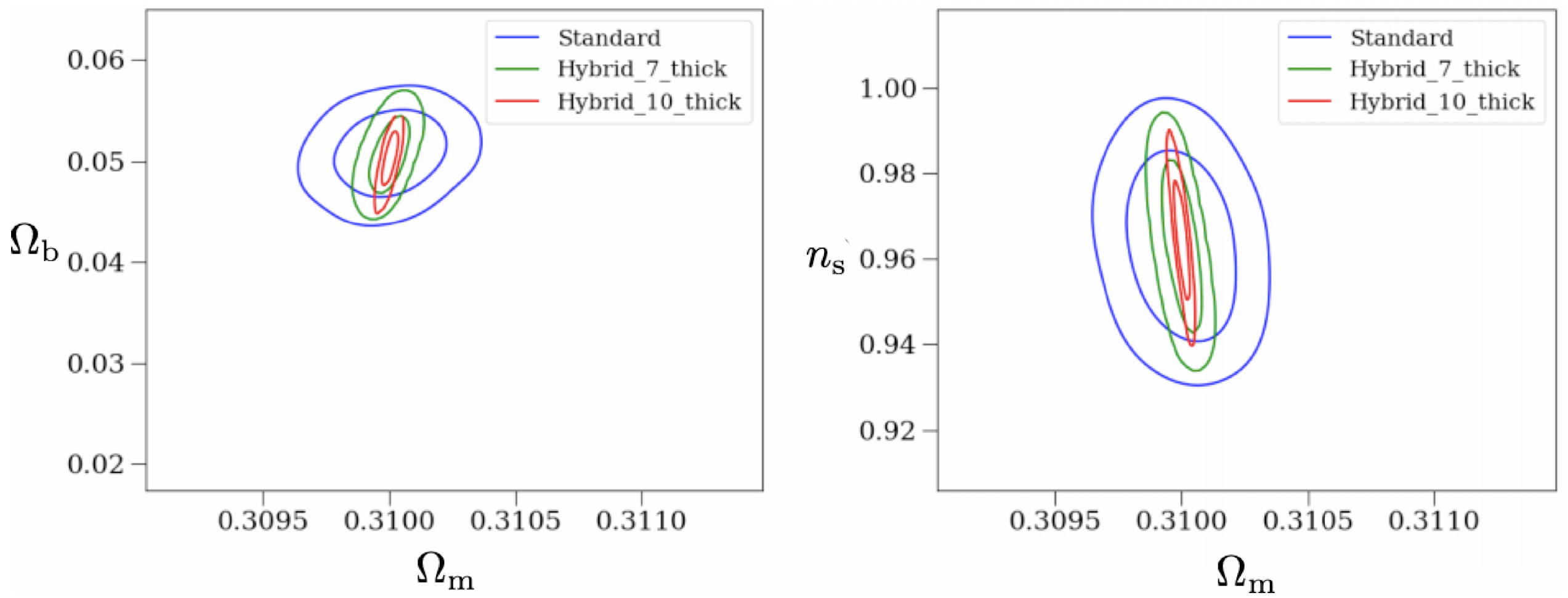}
\\
\hspace{-0.5cm}
\includegraphics[width=0.5\textwidth]{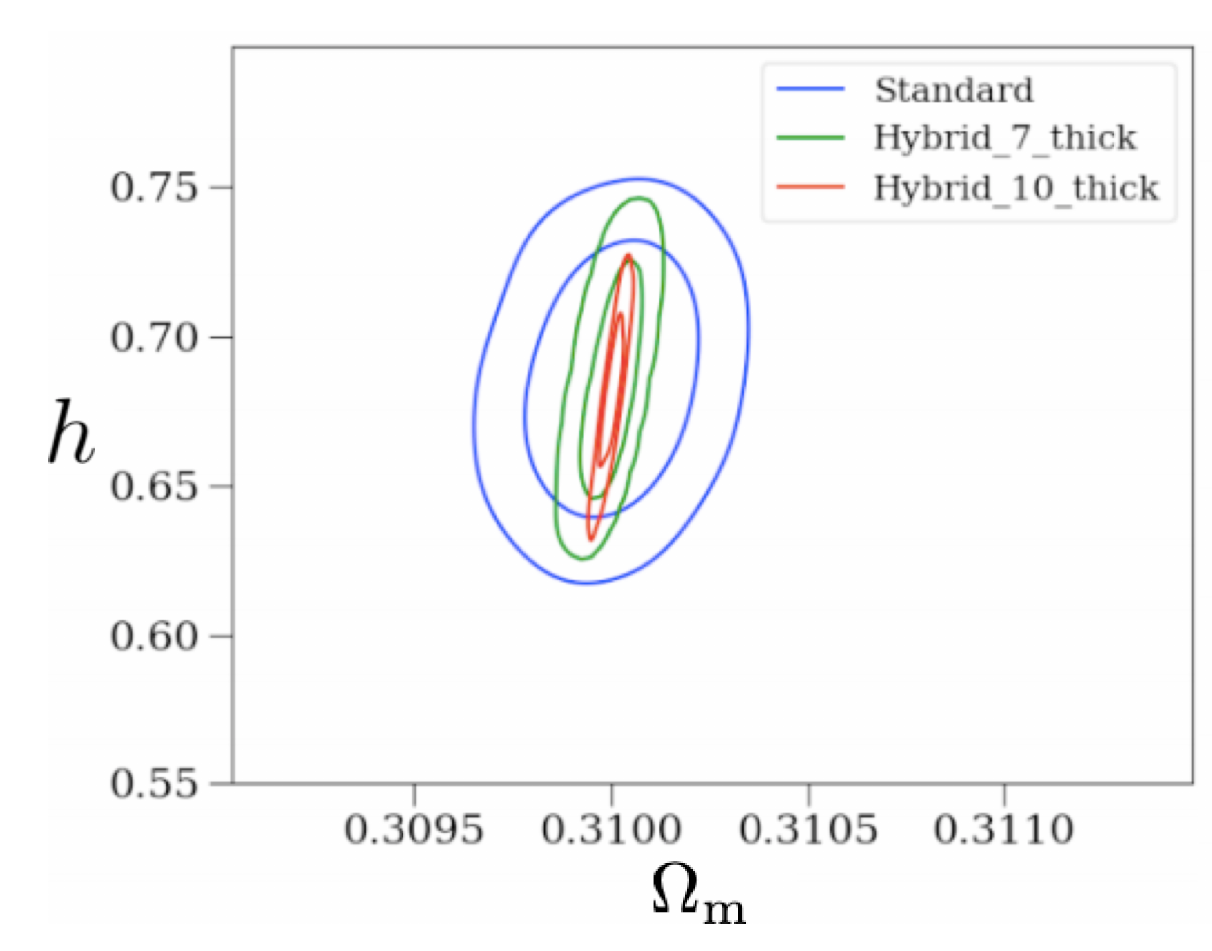}
\hspace*{0.4cm}
\includegraphics[width=0.47\textwidth,trim={0cm 0.56cm 0.6cm 0cm},clip]{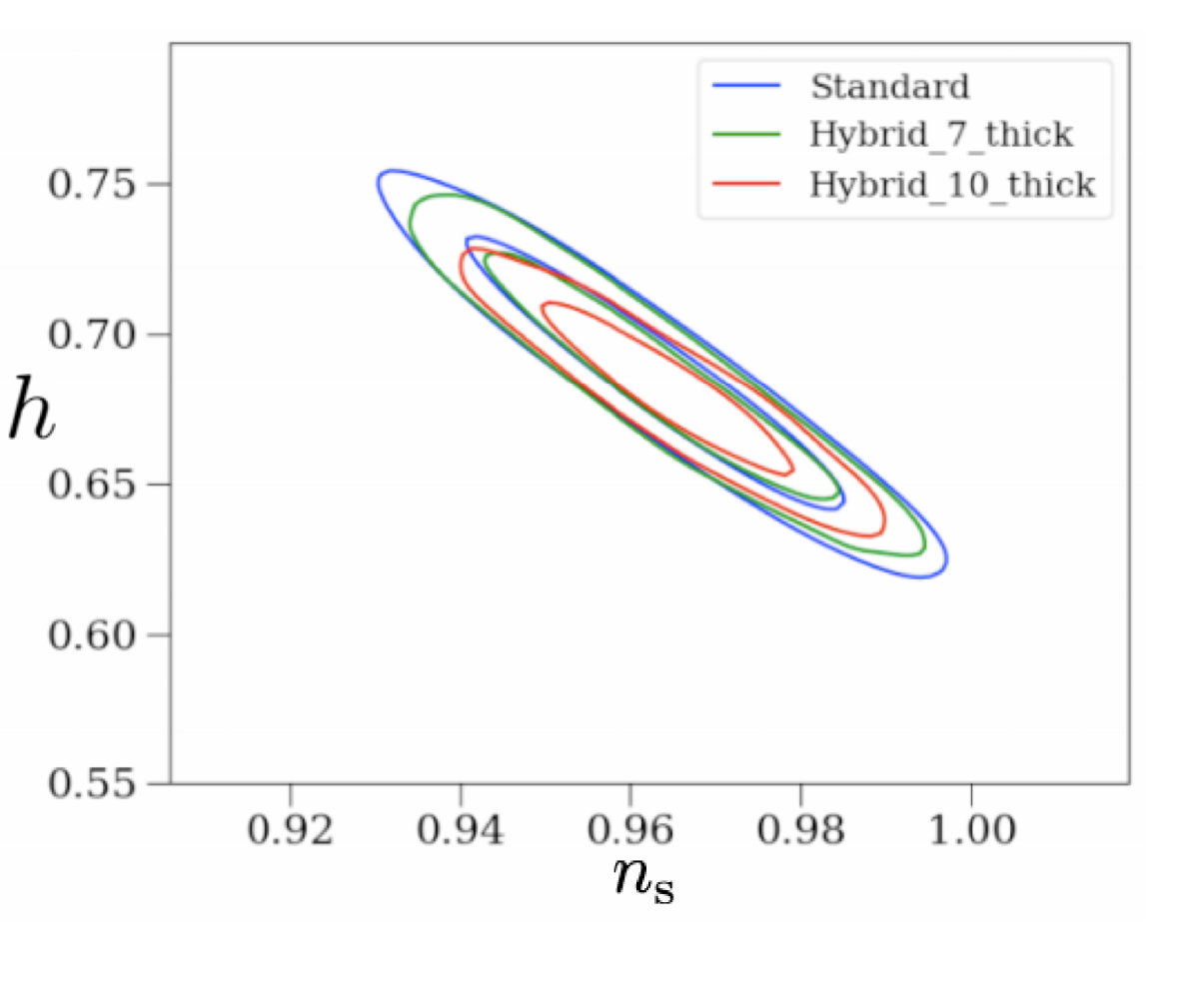}
%\vspace{-1.2cm}
%\hspace{0.7cm}
%\includegraphics[width=\textwidth,scale=0.5]{images/countour_2.pdf}
\caption{Two-dimensional joint posteriors on the following planes: $\Omega_{\rm m}-\Omega_{\rm b}$ (top left panel), $\Omega_{\rm m}-n_{\rm s}$ (top right panel),  $\Omega_{\rm m}-h$ (bottom left panel) and $n_{\rm s}-h$ (bottom right panel). Inner contours represent the 68 $\%$ confidence level areas, while the outer the 95 $\%$ areas.}
    \label{fig:All_three_trials_methods_1.pdf}
\end{figure}

To have a deeper understanding of the impact of the radial information retrieved by the hybrid approach, it is useful to look not only at the constraints on a single parameter, but rather at the cross-talks among different parameters, which tell us about intrinsic parameter degeneracies. \autoref{fig:All_three_trials_methods_1.pdf} shows the 68\% and 95\% joint marginal error contours on the two-dimensional parameter planes of the parameter set $\bm \theta$ for the three cases under investigation, i.e.\ the standard approach (blue contours), and the two binning configurations of the hybrid method (green and red contours for configuration 1 and 2 respectively). Looking at these plots it is evident that the new method is capable of constraining cosmological parameters better than the standard one, giving relative errors which are of the same order of magnitude but smaller. It is worth noting that the parameter $\Omega_{\rm m}$ is particularly better constrained with the hybrid procedure, having a relative error half of the one given by the standard method.
% 
% One would expect this trend for all the parameters. Nonetheless, the fact that we use only five $n_{\ell}$ multipole values, together with the fact that for the hybrid method we chose a wider binning than that of the original work by \citet{stef_specz}, could probably provide an explanation for the discrepancy between this work and the original article results. It is also worth reminding that here, for simplicity, we neglect both RSD and lensing magnification, and do not consider any nuisance parameter. In a future analysis these issue will have to be taken into account. However, the fact that the relative errors on all the parameters are found to be of the same order of magnitude with the original work constitutes a first achievement. We should also remark that the $\Omega_{\rm m}$ parameter is more strongly constrained, respecting the original work.

%\begin{figure}
%\centering 
%\includegraphics[width=\textwidth]{images/countour_2.pdf}
%\caption{Same as in  \autoref{fig:All_three_trials_methods_1.pdf} but for the $\Omega_{\rm m}-h$ plane (left panel) and the $n_{\rm s}-h$ plane (right panel).}
%    \label{fig:All_three_trials_methods_2.pdf}
%\end{figure}

\section{Conclusions}
\label{sec:conclusions}
In this work we make forecasts to compare the constraining power and reliability of a new hybrid tomographic method with the standard tomographic approach generally applied in the studies of spectroscopic galaxy clustering. We perform this comparison in a likelihood-based Bayesian approach going beyond the Fisher matrix analysis. We confirm that the standard and hybrid methods give comparable results, but the latter appears to be more constraining. On top of that, it saves computational time, as shown in \autoref{table:computational_times}. However, several approximations are made: we do not consider the RSD or correction due to lensing magnification (which nonetheless should be subdominant for fine redshift slicing). Also, we do not account for nuisance parameters, which are considered in the original paper. Finally, we work in the Limber approximation and calculate the angular power spectra at an exiguous number of multipole values to speed up the analysis computation. This, in principle, is not an issue, but in future works the hybrid method should be tested with finer binning, both in angular and in redshift space. Consequently, these approximations do not allow for a face-value comparison with the original paper results. In a forthcoming work we plan to reproduce the same analysis using finer binning, and introducing nuisance parameters as well as the contributions from RSD and magnification on the galaxy density field.

%\section*{Acknowledgements}
%We acknowledge support from the `Departments of Excellence 2018-2022' Grant (L.\ 232/2016) awarded by the Italian Ministry of Education, University and Research (\textsc{miur}). SC also acknowledges support by \textsc{miur} through Rita Levi Montalcini project `\textsc{prometheus} -- Probing and Relating Observables with Multi-wavelength Experiments To Help Enlightening the Universe's Structure' in the early stages of this project.

\section*{Author Contributions}
SC conceived the methodology. KT and GF created the algorithm. GF performed the analysis. SC and KT supervised the analysis. GF, KT, and SC wrote the article.

\section*{Funding Information}
The authors wish to thanks Roy Maartens and Jos\'e Fonseca for useful feedback on an early draft of this paper, as well as the anonymous referee who helped us improving the presentation of our results. The authors acknowledge support from the `Departments of Excellence 2018-2022' Grant (L.\ 232/2016) awarded by the Italian Ministry of Education, University and Research (\textsc{miur}). SC also acknowledges support from \textsc{miur} through Rita Levi Montalcini project `\textsc{prometheus} -- Probing and Relating Observables with Multi-wavelength Experiments To Help Enlightening the Universe's Structure' in the early stages of this project.

\section*{Conflicts of interest}
GF, KT, and SC declare none.

\section*{Data availability}
Data sharing not applicable – no new data generated.

%%%%%%%%%%%%%%%%%%%%%%%%%%%%%%%%%%%%%%%%%%%%%%%%%%



\bibliographystyle{mnras}
\bibliography{example}

\end{document}